\documentclass[10pt,twocolumn,twoside]{article}
\usepackage[a4paper, margin=1.5cm]{geometry}
\usepackage{amsmath,amssymb}
\usepackage{inputenc}
\usepackage[english]{babel}
\usepackage{graphicx}
\usepackage{wrapfig}
\usepackage{braket}
\usepackage{csquotes}
\usepackage[square,numbers]{natbib}
\RequirePackage{authblk}
\title{Lifetimes and quantum efficiencies of quantum dots deterministically positioned in photonic-crystal waveguides}

\author[1]{Xiao-Liu Chu}
\author[1]{Tommaso Pregnolato}
\author[2]{R\"udiger Schott}
\author[2]{Andreas D. Wieck}
\author[2]{Arne Ludwig}
\author[1]{Nir Rotenberg}
\author[1]{Peter Lodahl}

\affil[1]{Center for Hybrid Quantum Networks (Hy-Q), Niels Bohr Institute, University of Copenhagen, Blegdamsvej 17, DK-2100 Copenhagen, Denmark}
\affil[2]{Lehrstuhl f\"ur Angewandte Festk\"orperphysik, Ruhr-Universit\"at Bochum, Bochum, Germany}

\begin{document}
\maketitle





\begin{abstract}
Interfacing single emitters and photonic nanostructures enables modifying their emission properties, such as enhancing individual decay rates or controlling the emission direction. To achieve full control, the single emitter must be positioned in the nanostructures deterministically. Here, we use spectroscopy to gain spectral and spatial information about individual quantum dots in order to position each emitter in a pre-determined location in a unit cell of a photonic-crystal waveguide. Depending on the spatial and spectral positioning within the structured nanophotonic mode, we observe that the quantum dot emission can either be suppressed or enhanced. These results demonstrate the capacity of photonic-crystal waveguides to control the emission of single photons and that the ability to position quantum dots will be crucial to the creation of complex multi-emitter quantum photonic circuits.
\end{abstract}

\section{Introduction}
Photonic-crystal waveguides (PhCWs) control the flow of light and reshape the electromagnetic landscape at the nanoscale \cite{Novotny:12}, a powerful combination in the context of quantum optics.  By confining light to a small region, and slowing down its velocity, PhCWs enhance light-matter interactions enabling, for example, efficient photon-pair generation by spontaneous parametric downconversion \cite{Xiong:11}.  These same effects allow PhCWs to act as an efficient interface to solid-state quantum emitters such as quantum dots (QDs) \cite{Lodahl:2015} or defect centers in diamond \cite{Evans:18}.  In fact, PhCWs also strongly suppress emission into free-space \cite{Lodahl:2015,Javadi:18}, meaning that they can almost perfectly couple the emitters to guided photon modes \cite{Arcari:2014}.

\begin{figure}[h!]
	\centering
	\includegraphics[trim={0 0 0 0},clip, width=0.4\textwidth]{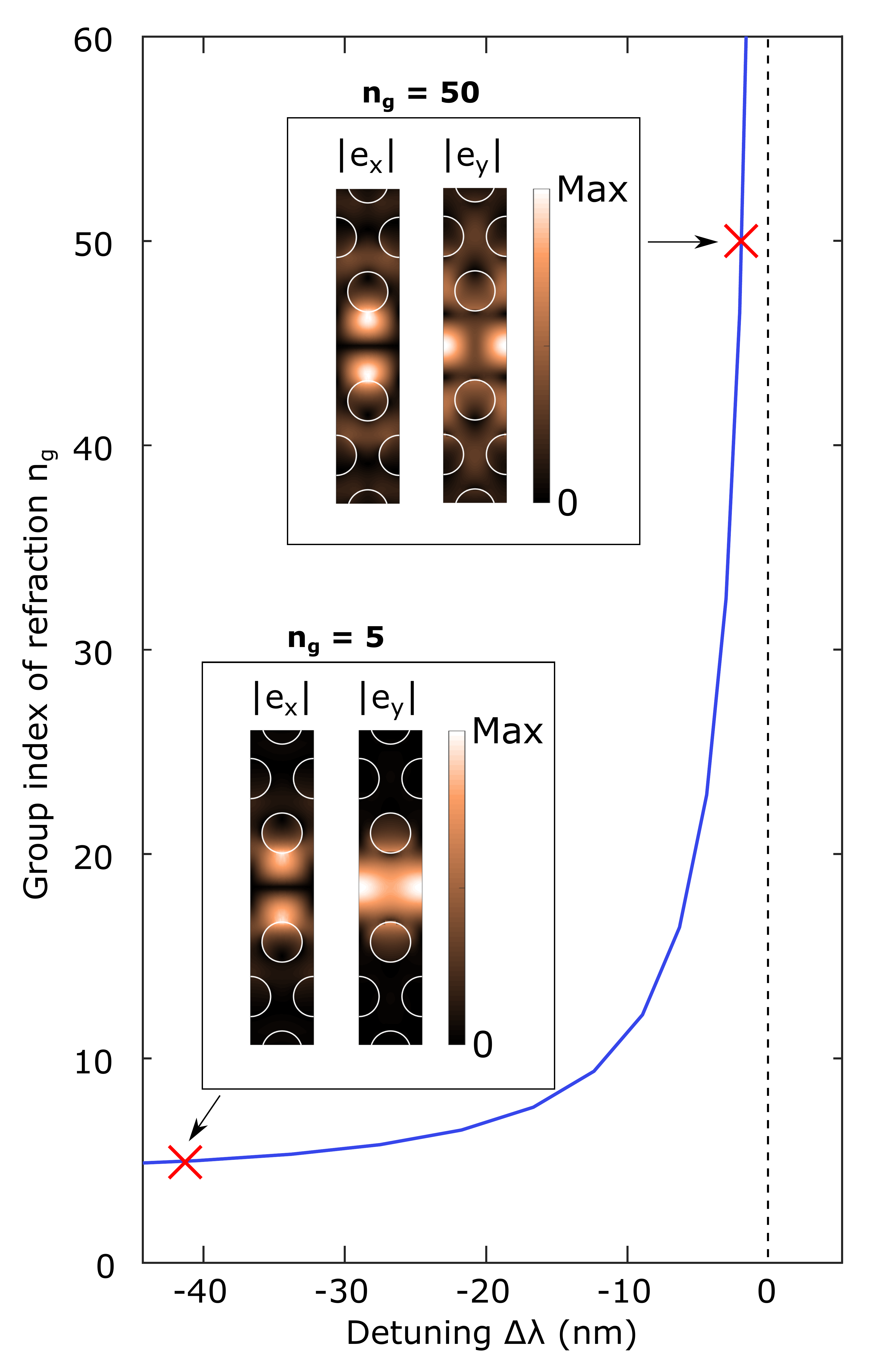}
	\caption{Position and wavelength dependent optical properties of a PhCW. The calculated group index $n_g$ diverges at the bandedge (detuning $\Delta \lambda = 0$), greatly enhancing light-matter interactions in the region. Insets: spatial map of the normalized amplitude of the in-plane electric field distributions for fast $\left(n_g = 5\right)$ and slow $\left(n_g = 50\right)$ light modes in a unit cell. The nanoscopic nature of the features in these light fields is evident as they are the same size as the holes of the PhCW (white circles, radius $r = 77$~nm).}
	\label{fig:PhCWProperties}
\end{figure}

The physics of emitter-photon coupling is captured by the Purcell factor, which quantifies the enhancement (or suppression) of emission into the PhCW $\Gamma_{\text{wg}}$ relative to the emission rate in bulk $\Gamma_{\text{B}}$ \cite{Purcell:46}. For PhCWs, the frequency $\omega$ and position $\mathbf{r}$ dependent Purcell factor can be written as \cite{Hughes:04}
\begin{equation}\label{eq:Fp}
F_{\text{P}}\left(\omega,\mathbf{r}\right)=\frac{3\pi ac^{2}}{\omega^{2}\sqrt{\varepsilon\left(\omega,\mathbf{r}\right)}}n_{g}\left(\omega\right)\left|\mathbf{d}^{*}\cdot\mathbf{e}\left(\omega,\mathbf{r}\right)\right|^{2},
\end{equation}
where $a$ is the lattice constant of the PhCW, $c$ is the speed of light in vacuum, and $\varepsilon$ is the dielectric constant of the material in which the emitter is embedded. The group index $n_g$ sets the velocity of the photonic mode according to $v_g=c/n_g$ and linearly enhances the emission rate, as shown in Eq.~\ref{eq:Fp}. Ideally, $n_g$ diverges at the photonic bandedge, as shown in the calculation presented in Fig.~\ref{fig:PhCWProperties}. In practise, unavoidable fabrication imperfections break the discrete symmetry of the photonic crystal, leading to unwanted scattering losses and capping the value of $n_g$ at $\approx 500$ \cite{Vlasov:05}.

Finally, the Purcell factor depends on the overlap between the transition dipole moment $\mathbf{d}$ (located at $\mathbf{r}$ and oscillating with angular frequency $\omega$) and the electric field of the photonic mode $\mathbf{e}\left(\omega,\mathbf{r}\right)$, which is normalised such that the integrated electrical energy over the unit cell is unity, at that position. The nanoscale geometry of PhCWs results in a complex and finely structured $\mathbf{e}\left(\omega,\mathbf{r}\right)$ that is known to contain regions of various possible field polarizations \cite{Rotenberg:15}. Examples of the in-plane field distributions $e_x\left(\omega,\mathbf{r}\right)$ and $e_y\left(\omega,\mathbf{r}\right)$, calculated for different $n_g$ modes, are shown in Fig.~\ref{fig:PhCWProperties}. Here, the structure of the light fields, with nanoscopic features, is readily visible.  

In total, the ability to engineer $n_g$ and $\mathbf{e}\left(\omega,\mathbf{r}\right)$ means it is possible to design PhCWs that enhance \cite{Thyrrestrup:10}, suppress \cite{Wang:2011} or even directionally couple emission \cite{Sollner:15,Coles:2016}, but only if the emitter can be precisely positioned.  High quality indium arsenide (InAs) QDs with excellent coherence properties \cite{Thyrrestrup:18} that produce indistinguishable photons with high count rates \cite{Ding:16,Kirsanske:17} demanded by emerging quantum technologies are normally grown via molecular-beam epitaxy using the Stranski-Krastanov method \cite{Lodahl:2015}. The growth process results in self-assembled and randomly distributed QDs, both spatially throughout each wafer, and in their emission wavelength. Unlocking the full power of the PhCW platform, as represented by Eq.~\ref{eq:Fp}, has therefore not been possible, and typical experiments involved the fabrication of many samples followed by a search for those that contain an emitter in a favorable position.

Here, we take the important step towards full control over the emission properties of QDs by deterministically placing these emitters at various predetermined locations within a PhCW unit cell, and near to the photonic bandgap. Spectroscopic and lifetime measurements of the QDs, both before and after nanofabrication, allow us to quantify the effect of the PhCW on the emission. We observe a spatial dependence of the emission properties and a significant improvement of the quantum efficiency of the QDs.  

\section{Deterministic integration of quantum dots in photonic-crystal waveguides}
\begin{figure*}[h!] 
	\centering
	\includegraphics[trim={0 0 0 0},clip, width=1\textwidth]{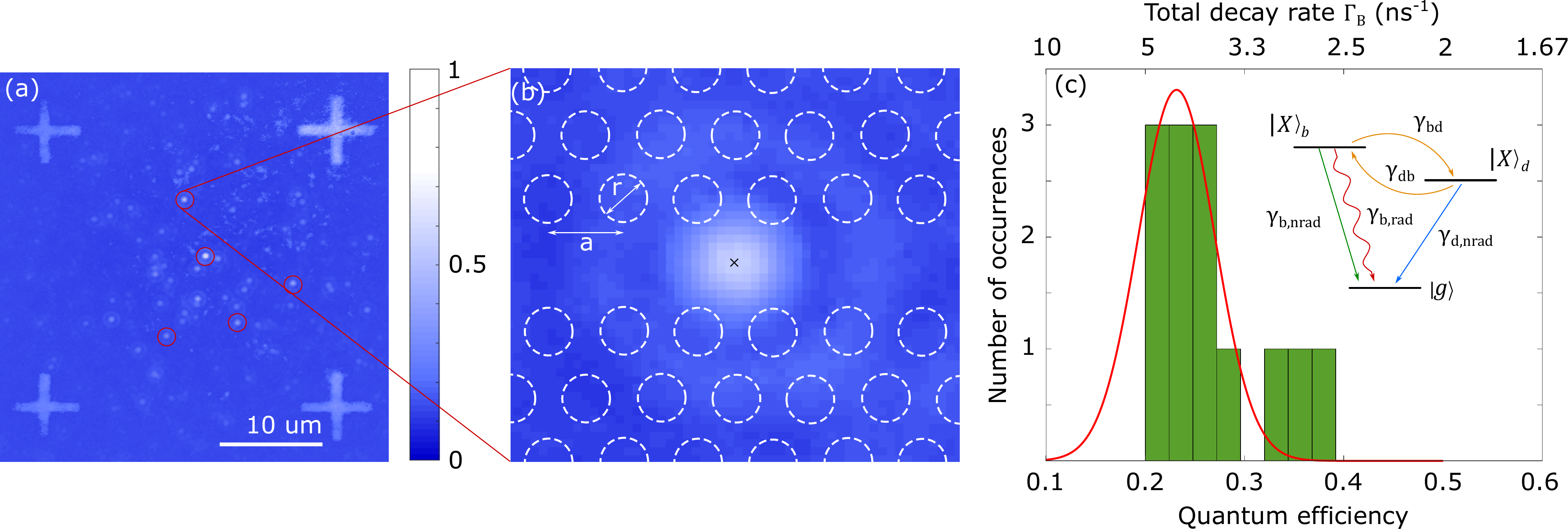}
	\caption{Localizing and characterizing QDs in bulk. (a) Superimposed camera image of the photoluminescence from QDs and from their host GaAs membrane.  Both the individual QDs and alignment markers are clearly visible in this figure, appearing as bright, diffraction limited spots and crosses, respectively.  Emission spectra from the brightest QDs (red circles) are taken. (b) This spectral and spatial information is used to design a PhCW about a QD, whose exact position is given by the black cross. The PhCW is defined by the radius $r$ of the holes and its lattice constant $a$. (c) Histogram of the quantum efficiency and decay times $\Gamma_{B}$ measured for QDs in bulk and a normal distribution fit to this data (red curve). Inset: a diagram of the energy levels for a neutral exciton, comprised of both a bright $\ket{X_b}$ and a dark $\ket{X_d}$ states with both radiative and nonradiative transitions.}
	\label{fig:Localization}
\end{figure*}

The first step of precise placement of QDs within PhCWs is to locate them and characterize their behaviour in bulk samples. The sample is a doped GaAs membrane that forms a p-i-n diode, which stabilizes and tunes the emission wavelength of the QDs \cite{Kirsanske:17}. To localize the QDs, we use the micro-photoluminesence protocol outlined in \cite{Pregnolato:19} and correlate images of QD emission to those of a global reference frame (here a grid of gold crosses, see Fig.~\ref{fig:Localization}a), to determine the absolute position of each QD. All measurements unless noted otherwise, are taken at a bias voltage of 0.3V.

Using this procedure, we localize each QD with an accuracy of 9.2 nm, similar to that of other micro-photoluminescence protocols \cite{Thon:09,Kojima:13,Sapienza:15,Liu:17,He:17} and resulting in a final QD-PhCW misalignment of only ($1\pm33$) nm \cite{Pregnolato:19}. We then measure the emission spectra of the brightest QDs and use this information to precisely design a PhCW around the QD, as shown in Fig.~\ref{fig:Localization}b. Here, the PhCW hole radii range from 71 to 76 nm and the lattice constant from 233 to 247 nm, nominally placing the bandedge within 10 nm of the emission wavelength (near 915 nm). 

Before nanofabrication, we carry out time-resolved measurements on the neutral excitons, $X^0$, of 13 QDs in bulk, which we determine from voltage spectra maps. We fit each lifetime measurement with a single exponential, presenting a histogram of the results in Fig.~\ref{fig:Localization}c (see Supplementary material for additional details on the QD states, lifetime measurements and fits). A normal distribution fit to this data (red curve) reveals a mean decay rate of 3.9~ns$^{-1}$ with a standard deviation of $\pm 0.7$~ns$^{-1}$, large compared to the lifetime in bulk semiconductor, which is about $1.1$~ns$^{-1}$ \cite{Stobbe:09}. To understand the nature of this large spread, we consider the energy-levels of a neutral exciton, which we sketch in the inset of Fig.~\ref{fig:Localization}(c). For an InAs QD, we expect two optically bright states, $\ket{X}_b$ and $\ket{Y}_b$, and two dark states, $\ket{X}_d$ and $\ket{Y}_d$, corresponding to the two in-plane transition dipole moments \cite{Wang:2011}. Each bright state can decay radiatively or non-radiatively to the ground state with a rate $\gamma_{b,r}$ or $\gamma_{b,nr}$, respectively, or transition to (or from) its dark state with a rate $\gamma_{bd}$ $\left(\gamma_{db}\right)$. In contrast, only non-radiative recombination with a rate $\gamma_{d,nr}$ is possible from the dark state. We neglect the coupling between the bright excitons, $\ket{X}_b$ $\rightleftarrows$ $\ket{Y}_b$, as the corresponding spin-flip processes are much slower than the other decay processes~\cite{Roszak:2007,Tsitsishvili:2010,Wang:2010}. 

The single-exponential nature of the measured lifetime implies that non-radiative processes are very significant (see Supplementary Materials). The fact that we do not observe contributions from both $\ket{X}_b$ and $\ket{Y}_b$ dipole transitions further supports the fact that the non-radiative rates dominate compared to the radiative ones. As such, the total decay rate is given by
\begin{equation}
\Gamma_{i} = \gamma_{i,r} + \gamma_{i,nr},\label{eq:Gamma_tot}
\end{equation}
where $i = B$ or $wg$, depending on whether the QD is embedded in bulk or in a waveguide, respectively, and all the non-radiative processes are included in $\gamma_{i,nr}$ such that the $b$ and $d$ subscripts referring to bright and dark states have been dropped. Alternatively, each QD can be characterized by its quantum efficiency,
\begin{equation}
QE_{i} = \frac{\gamma_{i,r}}{\gamma_{i,r}+\gamma_{i,nr}},\label{eq:QE}.
\end{equation}
which quantifies the probability that the QD will emit after excitation.  The radiative decay rate of InAs QDs in bulk GaAs, which emit near 900 nm, has been both calculated and measured to be $\gamma_{B,r} = 1$~ns$^{-1}$ \cite{Stobbe:09,Andersen:10}. Fig.~\ref{fig:Localization}c show that $QE_B$ ranges between 0.2 and 0.4, given this radiative rate. We note that the low $QE_B$ measured for these QDs is not the norm, as $QE_B$ close to unity has routinely been measured before \cite{Wang:2011}. 

\begin{figure}[!b] 
	\centering
	\includegraphics[width=0.4\textwidth]{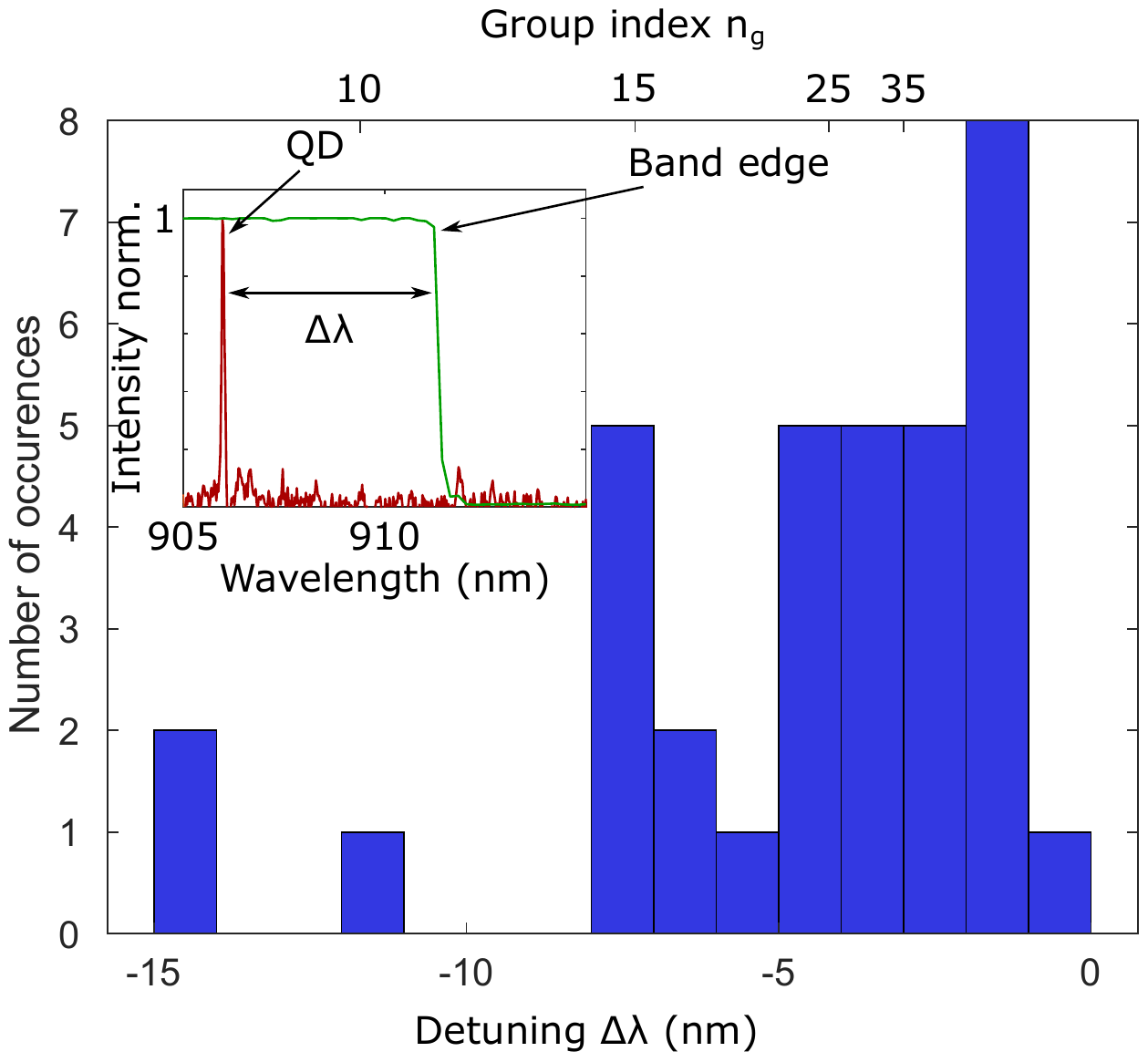}
	\caption{Spectral response of the QDs and the PhCWs. This is summarized in a histogram of the detuning $\Delta\lambda$ between the QD emission and the photonic band edge. Typical $n_g$ values for these detunings are given on the top axis. Inset: Example of a transmission curve and a QD emission spectrum, from which we calculate $\Delta\lambda$.}
	\label{fig:Spectra}
\end{figure}

Having located and characterized the QDs in bulk, we now fabricate 35 PhCWs with QDs equally distributed between 7 positions, as shown in Fig.~\ref{fig:Fabrication}. At these positions, $F_{\text{P}}$ for a transversely-oriented dipole $d_y$ varies by at maximum a factor of 30 (Fig.~\ref{fig:Fabrication}a). We show an example of the final, suspended PhCW with integrated grating couplers \cite{Pregnolato:19} in Fig.~\ref{fig:Fabrication}b. For each waveguide, micro-photoluminescence spectroscopy measurements (see inset to Fig.~\ref{fig:Spectra}, for an example) reveal whether or not a QD was successfully interfaced with the structure. From these, we find a yield of $94\%$ for structures with a nominal QD-surface separation $> 100$~nm, and $44\%$ for smaller separations. The latter positions are at $\left(0, 0.5a\right)$, $\left(0, 0.8a\right)$ and $\left(0, 1.3a\right)$, as shown in Fig.~\ref{fig:Fabrication}a, corresponding to nominal distances of 77.1, 48.6, and 30.7 nm from the nearest surface for $a=247$ nm and $r = 76$ nm. Moreover, it is likely that the effective separation is smaller due to a depletion region around the holes, where the QDs cannot be efficiently biased \cite{Berrier:07,Pregnolato:19}, meaning that the observed yield is consistent with our protocol accuracy.

We measure the transmission spectrum of each waveguide that contains a QD using a tunable laser, presenting an example in the inset to Fig.~\ref{fig:Spectra} (green curve).  This allows us to extract the detuning $\Delta\lambda$ between each QD and the band edge, where the transmission drops precipitously. A histogram of $\Delta\lambda$ for all QDs is shown in the main part of Fig.~\ref{fig:Spectra}, demonstrating that with few exceptions we are able to design and fabricate each individual PhCW such that the QDs lie within 10 nm of the band edge. As seen in Fig.~\ref{fig:PhCWProperties}, each $\Delta\lambda$ corresponds to a different $n_g$, and hence emission enhancement factor. Although the exact curvature of this function depends on the exact geometry of the PhCW, we give a rough indication of typical $n_g$'s for each separation in the top axis of Fig.~\ref{fig:Spectra}.

In total, these measurements demonstrate that we are able to precisely position QDs, both spatially and spectrally, with respect to PhCWs.

\section{Position dependent decay rate and quantum efficiency}
\begin{figure}[!b] 
	\centering
	\includegraphics[trim={0 0 0 0},clip, width=0.4\textwidth]{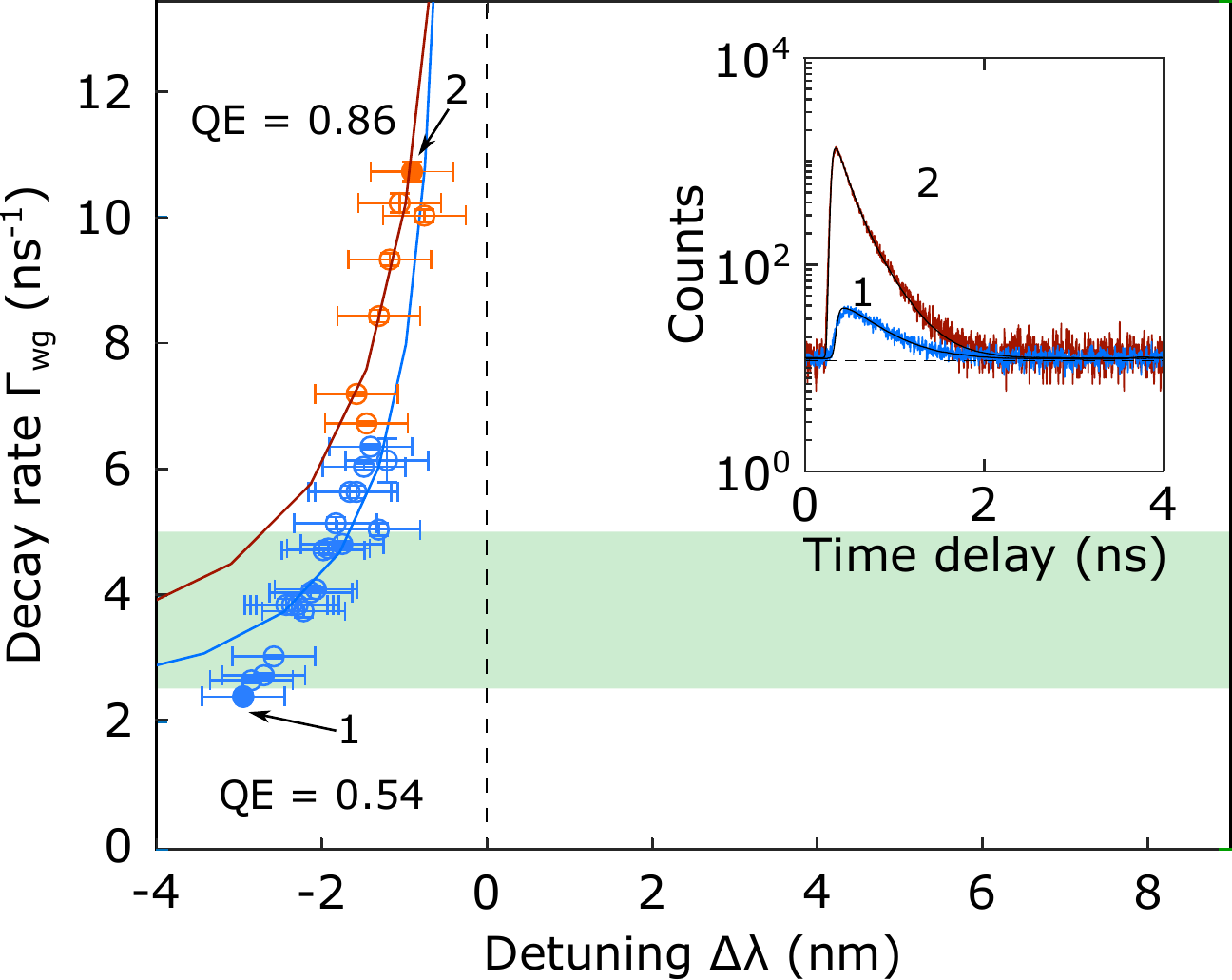}
	\caption{Photonic band-edge enhancement of $\Gamma_{wg}$. Lifetime measurements (symbols) reveal a sharp increase in the decay rate of two QDs as they are electrically tuned towards the band edge (dashed line), well above the bulk values (solid region).  The data are well-fitted using Eqs.~\ref{eq:Gamma_tot} and~\ref{eq:Gwgr} (solid curves).  Inset: examples of two lifetime measurements, corresponding to the two solid symbols in the main figure.}
	\label{fig:GammaVsdL}
\end{figure}

We can reformulate the radiative decay rate of QDs inside PhCW (Eq.~\ref{eq:Gamma_tot} using the Purcell factor (Eq.~\ref{eq:Fp}) as
\begin{align}
\gamma_{wg,r}\left(\omega,\mathbf{r}\right)& =F_{\text{P}}\left(\omega,\mathbf{r}\right)\gamma_{B,r} \approx A\left(\mathbf{r}\right) n_{g}\left(\omega\right)\gamma_{B,r},\label{eq:Gwgr}
\end{align}
where in the last line we assume that $A\left(\mathbf{r}\right)$, which is determined by the prefactor and the mode-dipole overlap, has only a weak wavelength dependence.  This assumption is reasonable in view of Fig.~\ref{fig:PhCWProperties}, where only minor changes to the electric field distributions were seen as $\Delta\lambda$ changed. We are therefore able to first study the effect of $\Delta\lambda$ on the emission, before proceeding to the more complex spatial dependence. 
\begin{figure*}[h] 
	\centering
	\includegraphics[trim={0 0 0 0},clip, width=0.9\textwidth]{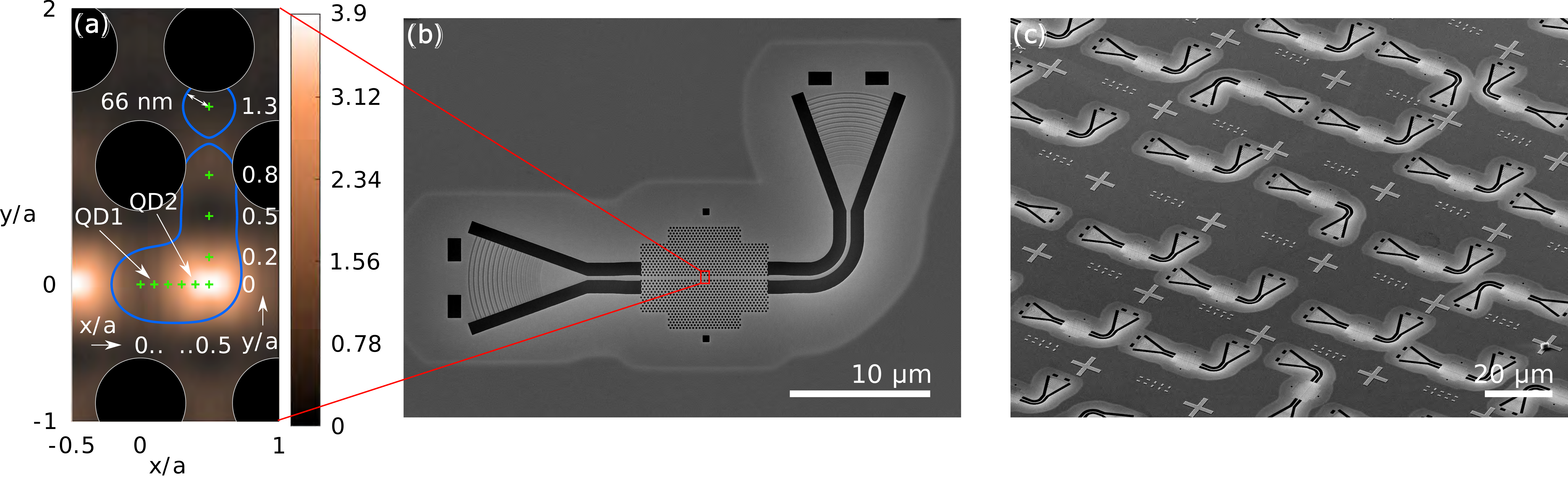}
	\caption{Deterministic integration of QDs into PhCWs (a) Designated QD positions (green crosses) within a unit cell of a PhCW, overlapped with the calculated $F_{\text{P}}$ for a $\hat{y}$-oriented transition dipole. Also shown are the holes of the PhCW (black circles) and the 66 nm alignment uncertainty of our protocol (blue curves). (b) SEM of a PhCW with integrated grating couplers. (c) A zoomed-out SEM shows many PhCWs, each of which contains a single, positioned QD. The randomness of the PhCW positions reflect that of the QDs.}
	\label{fig:Fabrication}
\end{figure*}
We begin by considering two emitters situated close to the photonic band edge, where the largest decay rate enhancement is expected. We consider two QDs, QD1 and QD2, located at $\left(0.2a, 0\right)$ and $\left(0.4a,0\right)$ in the unit cell and within 2 nm of the photonic band edge. Since QD2 lies closer to the mode maximum it should experience a slightly larger $F_{\text{P}}$ (c.f. Fig.~\ref{fig:Fabrication}a). At 0.3 V, the QDs emit at 915.1 nm and 904.7 nm respectively, which we change by scanning the external bias voltage between 0-0.6 V.  At each detuning, we measure and fit the decay rate, presenting the results for both QDs in Fig.~\ref{fig:GammaVsdL} (symbols). As expected, a significantly increased $\Gamma_{wg}$ is observed as the QD emission wavelength approaches the band edge, due to the increase in $n_g$.

Although both QDs in Fig.~\ref{fig:GammaVsdL} display the same wavelength-dependent trend, they differ in their details. Decay rates of QD1, for example, can be tuned between $2.4$ $ns^{-1}$ and $6.1$ $ns^{-1}$, meaning that it begins slower and ends up faster than the average $\Gamma_{B}$ (shaded region). This can be understood by considering that QD1 is located at position $\left(0.2a, 0\right)$, where we do not expect an enhancement of emission due to the Purcell factor $F_P$ (Fig.~\ref{fig:Fabrication}a). This suppression is then overcome through electrical tuning towards the band edge, and a corresponding increase of $n_g$ and $F_{\text{P}}$. In contrast, QD2 is positioned close to the field maximum, where its radiative decay rate is enhanced. We also calculate the $QE_{\text{wg}}$ of these two QDs, using the radiative and non-radiative decay rates extracted from the fits of Eqs.~\ref{eq:Gamma_tot} and~\ref{eq:Gwgr} to the data of Fig.~\ref{fig:GammaVsdL} (solid curves, see Supplementary Material for more information on the fitting process).  From these, we calculate that when QD1 is $\approx 2$~nm away from the band edge $QE_{\text{wg}} = 0.54$, which can be increased to 0.76 as $\Delta\lambda\rightarrow0$.  Similarly, for QD2 the $QE_{\text{wg}}$ increases from 0.78 to 0.86 as its emission wavelength approaches the band-edge. These results, and the agreement between our measurement and model, confirm that we observe strong wavelength-dependent slow-light emission enhancement \cite{Lund:2008,Thyrrestrup:10,Dewhurst:10}.

\begin{figure}[ht!] 
	\centering
	\includegraphics[trim={0 0 0 0},clip, width=0.45\textwidth]{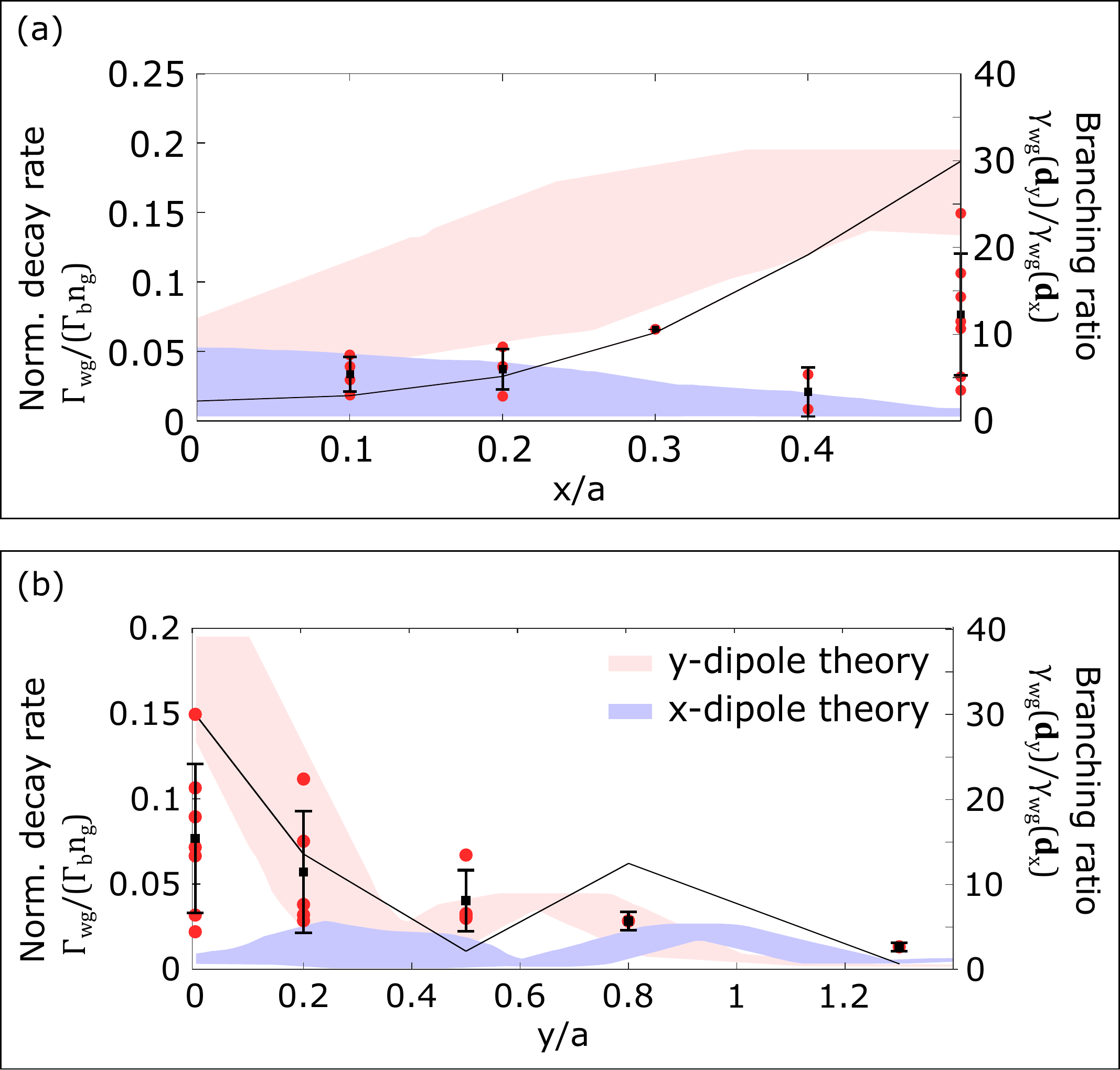}
	\caption{Normalized decay rate $\Gamma_{wg}/(n_g \Gamma_{B})$ of QDs located at different positions within a PhCW unit cell.  The QDs are located at (a) $y=0$ and shifted along $x$ or (b) $x=0.5a$ and shifted along $y$, as shown in Fig.~\ref{fig:Fabrication}a. Individual decay rates for each QD can be seen as red circles and the black squares are the mean values. The shaded regions correspond to the calculated normalized decay rates of the $\hat{x}$-oriented (blue) and $\hat{y}$-oriented (red) dipoles, and reflect the positional uncertainty of our protocol. The calculated branching ratio between the two dipoles, at each nominal position for $n_g=20$, is also given (black curve, right axis).}
	\label{fig:Position}
\end{figure}

We proceed to measure the decay rate $\Gamma_{\text{wg}}$ for the rest of the QDs in the PhCWs, and group them by their position in the unit cell. Note that in all cases, our measurements are well-fitted by single-exponential decay curves, which we interpret as due to the dominance of the $\hat{y}$-dipole at most positions. We normalize each data point to the $n_g$ experienced by that QD, and present the position dependent decay rates in Fig.~\ref{fig:Position}. 
Here, each individual measurement is represented by a red circle, while the mean and standard deviations are given by the black circles and error bars. For comparison, we also show the calculated normalized decay rate (at $n_g = 20$) for both transverse ($\hat{y}$-oriented) and longitudinal ($\hat{x}$-oriented) dipoles, using red and blue shaded regions that reflect the positional uncertainty of the QDs. We observe a qualitative agreement between our measurements and the theory, with the emission rate increasing as the $x$-coordinate of the QD increases (Fig.~\ref{fig:Position}a) and decreasing as the $y$-coordinate increases (Fig.~\ref{fig:Position}b). The differences between our measurements and theory are attributed to lack of statistics and the variable and large non-radiative decay rates of these QDs (c.f. Fig.~\ref{fig:PhCWProperties}c). The branching ratio $\gamma_{\text{wg,r}}\left(\hat{y}\text{-dipole}\right)/\gamma_{\text{wg,r}}\left(\hat{x}\text{-dipole}\right)$ for each nominal emitter position is also shown in Fig.~\ref{fig:Position} (black curve, right axis), demonstrating that the relative emission rates of the two bright excitons can be controlled through a careful positioning of the QD. A single exponential decay can be expected at the positions where the branching ratio is large, i.e. we primarily measure photons from one dipole transition. However, a deviation from a single exponential decay would be expected at positions where the branching ratio is small, such as position $0.1a$, since there is an equal contribution from two in-plane dipole transitions. The measured single exponential nature can therefore only be explained by a dominating nonradiative decay process. In fact, by considering the group of QDs in each position separately, we can extract information about their radiative and non-radiative decay rates and thus their $QE_{\text{wg}}$. 
\begin{figure}[ht!]
	\centering
	\includegraphics[trim={0 0 0 0},clip, width=0.4\textwidth]{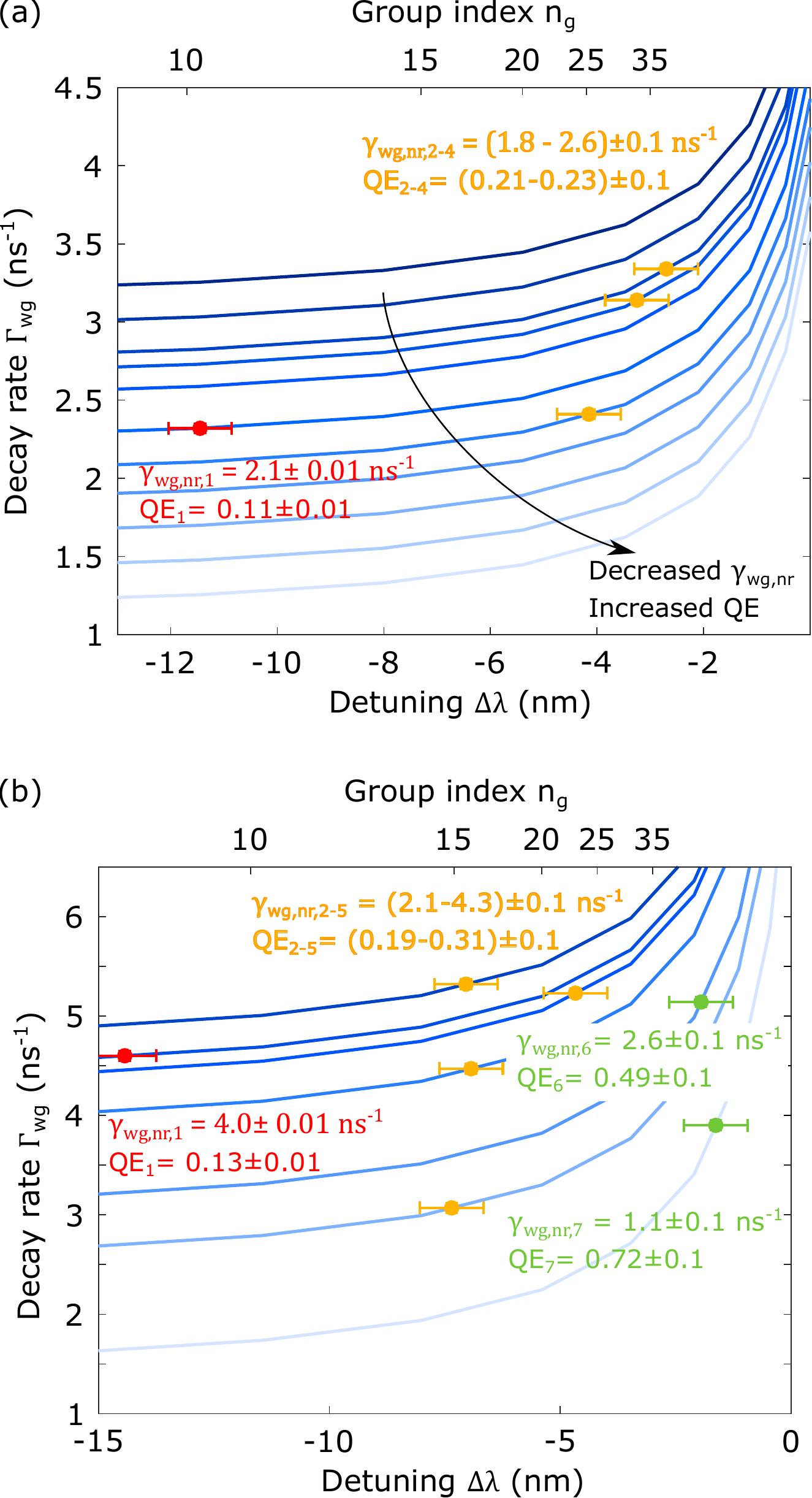}
	\caption{The decay rate $\Gamma_{\text{wg}}$ of QDs located at (a) $\left(0.5a, 0.5a\right)$ and (b) $\left(0.5a, 0\right)$ as a function of their respective detuning $\Delta\lambda$ (bottom axis) or respective $n_g$ (top axis). For each group, Eqs.~\ref{eq:Gamma_tot} and~\ref{eq:Gwgr} are used to fit the data, holding $\Gamma_{\text{wg}}/n_g$ constant for all QDs and allowing $\gamma_{\text{wg,nr}}$ to vary (values given in plot). From the fit parameters we calculate $QE_{\text{wg}}$ finding QDs whose emission is suppressed (red text), normal values for QDs in bulk (yellow text), or enhanced (green text), due to slow-light interactions.}
	\label{fig:QE_r}
\end{figure}
Two examples of this procedure are presented in Fig.~\ref{fig:QE_r}, in (a) for QDs located at $\left(0.5a, 0.5a\right)$ where we expect a strong suppression of emission, and in (b) at $\left(0.5a, 0\right)$ at the mode maximum (c.f. Fig.~\ref{fig:Fabrication}a).

We assume that all QDs at the same position have an identical $A\left(\mathbf{r}\right)$, and all the data for QDs at the same position can therefore be simultaneously modelled using Eqs.~\ref{eq:Gamma_tot} and~\ref{eq:Gwgr}, where only $n_g$ and the nonradiative rate $\gamma_{\text{wg,nr}}$ are allowed to vary.  In Fig.~\ref{fig:QE_r} we show the results of this fit, where the value of $\gamma_{\text{wg,nr}}$ that defines each curve is given. We extract both the radiative and non-radiative decay rates from the fits, which allow us to calculate $QE_{\text{wg}}$ of each QD using Eq.~\ref{eq:QE}. As expected, the radiative emission of the QD located at $\left(0.5a, 0.5a\right)$ (Fig.~\ref{fig:QE_r}a) and well away from the band edge is suppressed, resulting in a low $QE_{\text{wg}} = 0.11 \pm 0.01$.  Decreasing the detuning, and operating in the region where $n_g = 25 - 35$, compensates for this suppression.  For these QDs, we measure a $QE_{\text{wg}}$ that ranges between 0.21 and 0.23, commensurate with bulk $QE$ values (c.f. Fig.~\ref{fig:PhCWProperties}c), even if these QDs have larger $\gamma_{\text{wg,nr}}$. Here, we expect a low branching ratio and hence a biexponential decay, but do not observe it due to the large nonradiative rate $\gamma_{\text{wg,nr}}$. We observe a similar trend for QDs located at the mode maximum at $\left(0.5a, 0\right)$ (Fig.~\ref{fig:QE_r}b). Here, for a QD located well away from the band edge where $n_g = 8$, and which has a large $\gamma_{\text{wg,nr}}$, we find a low $QE_{\text{wg}} = 0.13$.  Increasing the $n_g \approx 15-20$ recovers bulk $QE_{\text{wg}}$, with values of 0.19-0.31 measured. Further increase of $n_g$ to beyond 25 results in a large enhancement of $\gamma_{\text{wg,r}}$, and measured $QE_{\text{wg}}$ values of 0.45 and 0.72, well beyond what we saw for bulk QDs.

\section{Conclusions}
In this article, we use a photoluminescence alignment procedure to experimentally study the temporal response and quantum efficiency of QDs deterministically distributed throughout a unit cell of a PhCW. We encounter a low QE due to the use of non-ideal QDs as already seen in bulk and show that emission can be further suppressed due to interaction with the highly structured photonic mode. This suppression can be overcome by exploiting the slow-light effect of the PhCW, resulting in enhanced QE. We use the position control to extract the QE, which we have previously done on QD ensembles \cite{Johansen:08}, and now extend to single QDs. In total, these results demonstrate the power of PhCWs to control the emission of single photons, and show that we can systematically and precisely unlock this ability by pre-determining the spectral and spatial position of emitters. Using this procedure in conjunction with emitters with QE = 1, we have the toolbox to map out the spatial dependence of the local density of states of PhCWs and other complex nanophotonic structures, as well as deterministically fabricate photon-emitter interfaces with near-unity coupling efficiencies. We anticipate that this capability will be critical to the creation of complex quantum photonic circuits \cite{Kimble:08,Brien:09}, as deterministic positioning is a key method for controllably scaling up nanophotonic systems and hence enable the coupling of multiple QDs.

\section*{Funding Information}
We gratefully acknowledge financial support from the Danish National Research Foundation (Center of Excellence “Hy-Q”), the Europe Research Council (ERC Advanced Grant “SCALE”), the
European Union’s Horizon 2020 research and innovation program
under the Marie Sk{\l}odowska Curie grant agreement no. 753067 (OPHOCS), Innovation Fund Denmark (Quantum Innovation Center “Qubiz”), and the Danish Research Infrastructure Grant (QUANTECH).

\section*{Acknowledgments}
The authors gratefully acknowledge Sandra {\O}. Madsen and Tim Schr\"oder for her help in the construction of the optical setup and Leonardo Midolo and Zhe Liu for assistance in fabrication. 

\section*{Supplemental Documents}
See Supplement 1 for supporting content.

\section*{Disclosure}
The authors declare no conflicts of interest.



 





\end{document}